\documentclass[superscriptaddress,twocolumn,aps,prl]{revtex4}
\usepackage{graphicx}% Include figure files
\begin{document}
\title{{\bf Theory of Phase Transition in the Evolutionary Minority Game}}
\author{Kan Chen}
\affiliation{Department of Computational Science, Faculty of Science, National University of Singapore, Singapore 117543}
\author{Bing-Hong Wang}
\affiliation{Department of Computational Science, Faculty of Science, National University of Singapore, Singapore 117543}
\affiliation{Department of Modern Physics, University of Science and Technology of China, Hefei, Anhui, 230026, China}
\author{Baosheng Yuan}
\affiliation{Department of Computational Science, Faculty of Science, National University of Singapore, Singapore 117543}

\date{\today }

\begin{abstract}

We discover the mechanism for the transition from self-segregation (into opposing groups) to clustering (towards cautious behaviors) in the evolutionary minority game (EMG). The mechanism is illustrated with a statistical mechanics analysis of a simplified EMG involving three groups of agents: two groups of opposing agents and one group of cautious agents. Two key factors affect the population distribution of the agents. One is the market impact (the self-interaction), which has been identified previously. The other is the market inefficiency due to the short-time imbalance in the number of agents using opposite strategies. Large market impact favors ``extreme" players who choose fixed strategies, while large market inefficiency favors cautious players. The phase transition depends on the number of agents ($N$), the reward-to-fine ratio ($R$), as well as the wealth reduction threshold ($d$) for switching strategy. When the rate for switching strategy is large, there is strong clustering of cautious agents. On the other hand, when $N$ is small, the market impact becomes large, and the extreme behavior is favored. 

\end{abstract}

\maketitle


Complex adaptive systems are ubiquitous in social, biological and economic sciences. In these systems agents adapt to the changes in the global environment, which are induced by the actions of the agents themselves. The main theme in the study of complex systems is to understand the emergent properties in the global dynamics. Of particular interest are the systems in which the agents have no direct interaction but compete to be in the minority; they modify their behaviors (strategies) based on the past experiences \cite{challet1,challet2}. Examples of such systems include financial markets \cite{challet3}, rush-hour traffic \cite{huberman}, and ecological systems. In the context of demand and supply in economic systems, the idea of the minority game is particularly relevant. If the demand is larger than the supply, the price of the goods will increase; this benefits the sellers who are in the minority. Many agent based models of economic systems and financial markets indeed incorporate the essence of the minority game.

In this letter we shall focus on the EMG proposed by Johnson, et al \cite{johnson}. The model is defined as follows. There are N (odd number) agents. At each round they choose to enter Room 0 (sell a stock or choose route A) or Room 1 (buy a stock or choose route B). At the end of each round the agents in the room with fewer agents (in the minority) win a point; while the agents in the room with more agents (in the majority) lose a point. The winning room numbers (0 or 1) are recorded, and they form a historical record of the game. All agents share the common memory  containing the outcomes from the most recent occurrences of all $2^m$ possible bit strings of length $m$. The basic strategy is derived from the common memory and is changing dynamically. Given the current $m-bit$ string, the basic strategy is simply to choose the winning room number after the most recent pattern of same $m-bit$ string in the historical record. To use the basic strategy is thus to follow the trend. In the EMG each agent is assigned a probability $p$: he will adopt the basic strategy with probability $p$ and adopt the opposite of the basic strategy with probability $1-p$. The agents with $p=0$ or $p=1$ are ``extreme'' players, while the agents with $p=1/2$ are cautious players. The game and its outcomes evolve as less successful agents attempt to modify their $p$ values. This is achieved by allowing the agents with the accumulated wealth less than $d$ ($d<0$) to change their $p$ values. In the original EMG model, the  new p value is chosen randomly in the interval of width $\Delta p$ centered around its original $p$ value. His wealth is reset to zero and the game continues. Thus in the EMG the agents constantly learn from mistakes and adapt their strategies as the game evolves.

A remarkable feature emerges from the study of the EMG: the agents self-segregate into two opposing extreme groups with $p\sim 0$ and $p\sim 1$ \cite{johnson,lo1,lo2}. This conclusion is rather robust; it does not depend on $N$, $d$, $\Delta p$, $m$, or the initial distribution of $p$. The final distribution always has symmetric U-shape. This leads to the following conclusion: in order to succeed in a completive society the agent must take extreme positions (either always follows a basic strategy or goes against it). This behavior can be explained by the market impact of the agents' own actions which largely penalizes the cautious agents \cite{lo2}. By introducing the reward-to-fine ratio $R$, Hod and Nakar found that the above conclusion is only robust when $R \ge 1$. When $R<1$ there is a tendency to cluster towards cautious behaviors and the distribution of the $p$ value, $P(p)$, may evolve to an inverted-U shape with the peak at the middle. In some ranges of the parameters M-shape distributions are also observed. To explain the clustering of cautious agents, Hod gives a phenomenological theory relating the accumulated wealth reduction to a random walk with time-dependent oscillating probabilities \cite{hod3}. However, the dynamical mechanism for the phase transition is not clear. This letter aims to present such mechanism from the analysis based on statistical mechanics.

\begin{figure}
\includegraphics*[width=8.5cm]{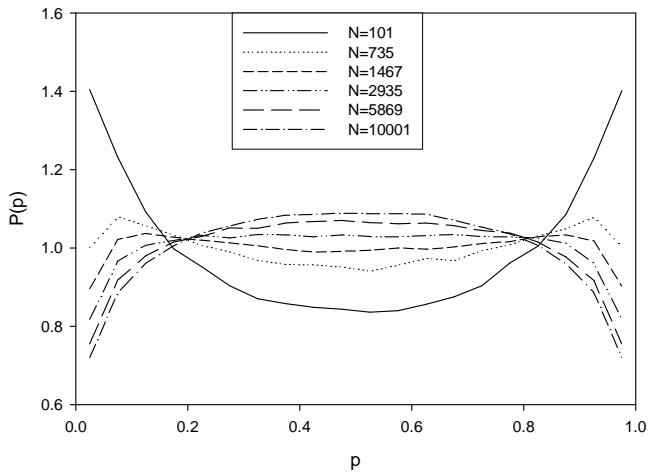}
\caption{ The distribution P(p) for R=0.971 and $d=-4$. A set of values of $N = 101, 735, 1467, 2935, 5869$, and $10001$ are used. The distribution is obtained by averaging over 100,000 time steps and 10 independent runs}
\end{figure}

We first present our numerical results which show that the transition from self-segregation to clustering is generic for $R<1$. We have performed extensive simulations of EMG for a wide range of the values of the parameters, $N, R$, and $d$. The transition depends on all three parameters, $N$, $R$, and $d$. Figure 1 shows the distribution $P(p)$ for $R=0.971$, $d=-4$, and $N=101, 735, 1467, 2935, 5869$ and $10001$. For a given $R$ ($<1$) and $d$, we observe the transition from self-segregation to clustering as the number of agents $N$ increases. The shape of the distribution $P(p)$ changes from a U-shape to an inverted U shape (near the transition point $P(p)$ has M-shape). The standard deviation $\sigma_p$ of the distribution decreases as $N$ increases. We define the critical value $N_c$ as the value of $N$ when $\sigma_p$ equal to the standard deviation of the uniform distribution, i.e. when $\sigma_p^2 = \int^1_0 (p-1/2)^2P(p)dp$
equal to $1/12$. Our results can be summarized by the general expression for the critical value $N_c =\left[\frac{|d|}{A(1-R)}\right]^2$,
where $A$ is a constant of order one. Alternatively one might view the transition by varying $d$ with fixed $N$ and $R$. As $|d|$ increases the system changes from clustering to self-segregation. The critical value is then given by $|d_c| = A(1-R)\sqrt{N}$. Figure 2 plots $N_c$ vs $|d|$ for various $R$. When $R\rightarrow 1$ the clustering only occurs for very large $N$ or very small $|d|$. At $R=1$ the clustering disappears and the behavior of self-segregation becomes robust.

\begin{figure}
\includegraphics*[width=8.5cm]{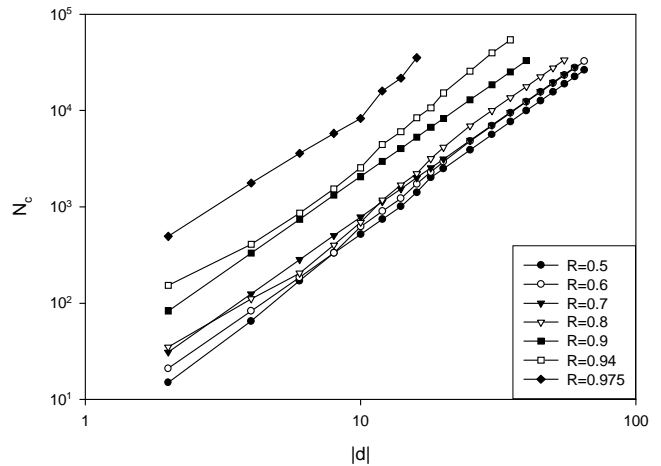}
\caption{ The critical value $|d_c|$ vs $N$ for $R=0.5,0.6,0.7,0.8,0.9,0.94$, and $0.975$.}
\end{figure}

Hod and Nakar explain that $R<1$ corresponds to  difficult situations (tough environments) in which the agents tend to be confused and indecisive and thus become cautious. We find that the rate of strategy switching (which depends on both $R$ and $d$) affect the distribution of the agents more directly. For $R<1$ the agent switches its strategy every $2|d|/(1-R)$ time steps on average. So when $R$ or $|d|$ is small, the agents have less patience and switch their strategies more frequently; this, as we explain below, causes large market inefficiency and thus favors cautious agents. It is the rapid adaptation that makes the agents ``confused'' and ``indecisive''.  On the other hand, when the number of agents is small, the market impact becomes large. Take for example a population consists of only three agents with $p=0, 1/2$, and 1 respectively. The cautious agent (with $p=1/2$) always loses because he is always in the majority, while the extreme agents are in the majority half of the times. In this case the cautious agent experiences the full market impact of his own action. Indeed our data show that when $N$ is small enough the self-segregation into extreme behaviors dominates.

We now show that the mechanism for clustering around $p=1/2$ and the transition to self-segregation can be understood from a simplified model in which $p$ takes only three possible values $p=0,1/2$, and $1$. The agents in Group 0 (with $p=0$) makes the opposite decision from the agents in Group 1 (with $p=1$). We denotes the group with $p=1/2$ ``Group m''. The probability of winning only depends on $N_0, N_m, N_1$, which are the respective numbers of agents in Group 0, m, and 1.

We begin by evaluating the average wealth reduction for the agents in each of the three groups. 
Let $n$ be the number of agents in Group m making the same decision (let us call it decision A) as those in Group 0 ($N_m - n$ will then be the number of agents in Group m making the same decision (decision B) as those in Group 1). If $N_0 + n < (N_m - n) + N_1$, or $n< N_m/2 + (N_1 - N_0)/2$, the agents making decision A will win; when $n> N_m/2 + (N_1 - N_0)/2$, the agents making decision B will win. The winner has its wealth increased by $R$, while the loser has its wealth reduced by 1. With $N_0$, $N_m$, and $N_1$ fixed, the probability of winning depends on $n$. 

When $N_m \gg 1$, the distribution of $n$ can be approximated by a Gaussian distribution $P(n) = \frac{1}{\sqrt{2\pi}\sigma_m}\exp(-(n-N_m/2)^2/(2\sigma_m^2)),$
where $\sigma_m = \sqrt{N_m}/2$. Given the distribution, one can write down the average wealth change for the agents in Group 0,
$$\Delta w_0 = R\int^{N_m/2+N_d/2}_0 P(n)dn - \int^{N_m}_{N_m/2+N_d/2}P(n)dn,$$
where $N_d = N_1 - N_0$. This can be rewritten in term of the error function $\text{erf}(x) = \frac{2}{\sqrt{\pi}}\int^x_0e^{-t^2}dt$,
\begin{equation}
\Delta w_0  = -\frac{1-R}{2} + \frac{1+R}{2}\text{erf}\left(\frac{N_d}{2\sqrt{2}\sigma_m}\right).
\end{equation}
Similarly we can derive the average wealth change for the agents in Group 1,
\begin{equation}
\Delta w_1  = -\frac{1-R}{2} - \frac{1+R}{2}\text{erf}\left(\frac{N_d}{2\sqrt{2}\sigma_m}\right)
\end{equation}

Since the number $N_0$ and $N_1$ are fluctuating, and on the average $N_0$ and $N_1$ should be the same, we can average out the short time fluctuations in $N_d$. This allows us to find out how the agents in the ``extreme'' groups compare with the cautious agents in Group m in the long run. The average wealth change of the agents in Group 0 and 1 is given by $\Delta w_e= (N_0  \Delta w_0 + N_1 \Delta w_1)/(N_0 + N_1)$. Substituting the expressions for $\Delta w_0$ and $\Delta w_1$, we have
\begin{equation}
\Delta w_e =-\frac{1-R}{2} - \frac{1+R}{2}\frac{N_d}{N_0 + N_1}\text{erf}\left(\frac{N_d}{2\sqrt{2}\sigma_m}\right)
\end{equation}

Note that the second term in $\Delta w_e$, which is due to the fluctuations in $N_d$, is always negative (since $\text{erf}(x)$ is an odd function). When $N_0 \neq N_1$, the winning probabilities for making decision A and decision B are not equal, and the market is not efficient. Thus this term can be interpreted as the cost due to market inefficiency. Large market inefficiency on average penalize the players taking ``extreme" positions more.

For the agents in Group m, if $n< N_m/2 + N_d/2$, then $n$ agents in the group win, while $N_m -n $ agents in the group lose. On the other hand, if $n> N_m/2 + N_d/2$, then $N_m -n$ agents in the group win, but $n$ agents lose. We need to take these two cases into account when evaluating the average.
$$\Delta w_m = \frac{1}{N_m}\left[ \int^{N_m/2+N_d/2}_0(Rn-(N_m-n)) P(n)dn\right.$$
$$\;\; \left.+\int^{N_m}_{N_m/2+N_d/2}(R(N_m -n)-n)P(n)dn\right].$$
After a few algebraic steps, we arrive at
\begin{equation}
\Delta w_m = -(1-R)/2 -\frac{1+R}{\sqrt{2\pi N_m}}  \exp(-N_d^2/(2N_m))
\end{equation}

The first term in $\Delta_m$ is the same as that in $\Delta_e$. The second term can be interpreted as the market impact \cite{lo2}. The magnitude of the term is in fact the largest when $N_d = 0$. Large market impact (self-interaction) penalizes the cautious players; their own decisions increase their chances of being in the majority and hence their chances of losing.

To determine the transition from clustering to self-segregation, we need to calculate the distribution of $N_d$ which allows us to evaluate $\Delta w_e$ and $\Delta w_m$. Let us denote the change in $N_d$ in one time step as $\delta N$. On average $\delta N = 2 N_0/(|d|/((1-R)/2))=N_0(1-R)/|d|$; this is the average number of extreme agents switching their strategies per time step (adaptation rate). The factor 2 is included because the agent only loses about half of the times. $|d|/((1-R)/2)$ is the average time step taken before the wealth threshold is reached. The dynamics of $N_d$ can be described as a random walk with mean reversal (there is a higher probability moving towards $N_d = 0$). The individual step of the walk is given by $\pm \delta N$. The probability for changing from $N_d$ to $N_d + \delta N$ is given by $W_{+}(N_d)$,  and the probability for changing to $N_d -\delta N$ is given by $W_{-}$, where $W_{\pm} = \frac{1}{2}[1 \mp \text{erf}(N_d/(2\sqrt{2}\sigma_m)]$. The steady state probability distribution $Q(N_d)$ for $N_d$ should satisfy
\begin{eqnarray}
Q(N_d) &=& W_{-}(N_d +\delta N)Q(N_d + \delta N) \nonumber \\
	& &+W_{+}(N_d - \delta N))Q(N_d - \delta N).
\end{eqnarray}
For small $\delta N$ one can convert the above equation to a differential equation. The solution of $Q(N_d)$ is given by
\begin{equation}
Q(N_d) \propto \exp(-\frac{2}{\delta N}\int^{N_d}_0 \text{erf}(\frac{n}{2\sqrt{2}\sigma_m})dn)
\end{equation}
Now we average $\Delta w_e$ and $\Delta w_m$ over the distribution of $Q(N_d)$. We can easily obtain that
\begin{equation}
\Delta w_e = -\frac{1-R}{2}-\frac{(1+R)}{2}\frac{\delta N}{2(N_0 + N_1)}
\end{equation}
$\Delta w_m$, on the other hand, is given by 
$$\Delta w_m = -(1-R)/2 -\frac{1+R}{\sqrt{2\pi N_m}} < \exp(-N_d^2/(2N_m))>,$$
where the average is over the distribution $Q(N_d)$. This can be approximated as
$$\Delta w_m  \sim  -\frac{1-R}{2}-\frac{1+R}{\sqrt{2\pi}}\frac{1}{\sqrt{N_m +\sigma_d^2}}, $$
since in the range  $N_d < \sigma_m$, where the main contribution to the average comes from, $Q(N_d)$ can be well approximated by a Gaussian distribution centered at zero with width $\sigma_d = \sqrt{\frac{\sqrt{2\pi}}{2}}\sqrt{\sigma_m \delta N}$. 
At the critical point, $N_0=N_1=N_m = N/3$, and $\Delta w_e = \Delta w_m$. It is easy to verify that this occurs when $\delta N \sim \sqrt{N_m}$.  As $\delta N =N_0(1-R)/|d|$, the crossover value for $|d|$ is $|d_c| = A_0  (1-R)\sqrt{N}$, where $A_0$ is a constant of the order one. 

In the above derivation we simply use the averaged value for $\delta N$. This underestimates the magnitude of $\Delta w_e$. For $R$ close to 1, the strategy switching in the ``extreme" group is rather intermittent. There are no agents switching strategy for many time steps, but in a single step many agents in the group switch strategies. A loss at a single round, for example, will not make the agents in the extreme group to switch strategy if they had won in the previous two rounds. We can take this intermittency into account, by introducing the probability $z$ that strategy switching occurs in the extreme group after it loses. We leave out the case $\delta N = 0$, since it does not affect the distribution of $N_d$. The average  $\delta N$ is now
 $N_0(1-R)/(z|d|)$. The crossover value for $d$ is then given by $|d_c| = (A_0/z) (1-R)\sqrt{N} \equiv A(1-R)\sqrt{N}$.  If $\delta N$ is close to its averaged value and $z\sim 1$, $A$ is of the order one. The broader the distribution of $\delta N$ and the larger the intermittency in strategy switching among the agents in the extreme groups, the larger the value of $A$. One can estimate the upper bound for $A$ as follows.  The probability $z$ and $\delta N$ are related to the wealth distribution of the agents in the extreme groups. The minimum width of the wealth distribution is $|d|$, so $\delta N < N/|d|$. The upper bound in $d_c$ is thus obtained with $\delta N = N/d$ and $z=1-R$; this leads to $d_c \sim \sqrt{N}$, or $A \sim 1/(1-R)$. Figure 3 shows $A$ vs N for various $R$ values. One can see that $A$ becomes independent of $N$ for sufficiently large $N$ (this means that $|d_c| \propto \sqrt{N}$ holds well numerically). The value of $A$ indeed approaches the upper bound $A_0/(1-R)$ for the three-group EMG when $N > 1/(1-R)^2$. This can be understood by the following simple argument: The width of the wealth distribution is close to $|d|$ when $|d|$ is greater than the wealth fluctuation, which is roughly 
$\sqrt{d/(1-R)}$,
 given that the average time for strategy switching is about $|d|/(1-R)$. Thus when $|d| > 1/(1-R)$ or $N > 1/(1-R)^2$, the upper bound for $A$ is reached. However, this is likely to be the unique feature for the three-group EMG model. For the original EMG the value of $A$ is of order one for a wide range of $R$, as can also be seen from Figure 3.

\begin{figure}
\includegraphics*[width=8.5cm]{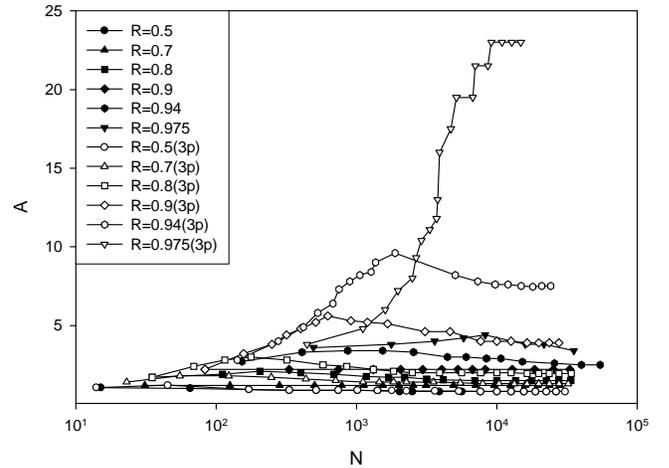}
\caption{ $A$ vs $N$ for various values of $R$. The results from the three-group EMG and the original EMG with random redistribution are shown}
\end{figure}

The theory can be generalized to the original EMG model by generalizing the definition of $N_d$ to $N_d =2(\bar{p}-1/2)$, where $\bar{p}$ is the average of the p values among all the agents at a given time step. The market inefficiency is again measured by the fluctuation in $N_d$. Consider the version in which the agent choose a new $p$ randomly when its wealth is below $d$, then we can argue that $\delta N$ (the average change in $N_d$) is again given by $\delta N \sim N(1-R)/|d|$. So we have $|d_c| = A_0(1-R)\sqrt{N}$; this works well because the fluctuation of $\delta N$ is likely to be much smaller in the original model than in the three-group model. We can also understand the version of the model in which the new $p$ value is chosen in the interval of width $\delta p$ around the old $p$ value. Since a smaller $\delta p$ leads to a smaller $\delta N$, the cost due to market inefficiency is reduced. This favors the ``extreme" agents ($|d_c|$ is smaller for a smaller $\delta p$); it is consistent with the results obtained in Ref. \cite{burgos}. Ref.~\cite{hod2} found that the periodic boundary condition used in the redistribution of the $p$ value favors clustering. This is also not surprising. When the boundary condition is periodic in $p$, $\delta N$ is effectively increased, because some $p=0$ agents can switch to $p=1$ agents, even when $\delta p$ is small. 

In conclusion, we have derived a general formalism for studying the transition from clustering to self-segregation based on the statistical mechanics of a simplified three-group model. We find that frequent strategy switching leads to market inefficiency which favors the clustering of cautious agents. A general expression relating the number of agents, the wealth threshold, and the reward-to-fine ratio at the critical point is derived. This expression is found to be equally valid for the general EMG.

This work is supported by the National University of Singapore through the research grant R-151-000-028-112. BHW also acknowledges the support by the Special Funds for National Basic Research Major Project "Nonlinear Science", by the National Natural Science Foundation of China (Nos.19932020,
19974039 and 70271070), and by the China-Canada University Industry
Partnership Program (CCUIPP-NSFC Grant No. 70142005).

\end{document}